# $LaCo_2B_2$: A Co-based layered superconductor with a $ThCr_2Si_2$–type structure

Hiroshi Mizoguchi,[1] Toshiaki Kuroda,[2] Toshio Kamiya,[2] and Hideo Hosono[1,2]*

[1]Frontier Research Center, Tokyo Institute of Technology, 4259 Nagatsuta, Midori-ku, Yokohama 226-8503, Japan

[2]Materials and Structures Laboratory, Tokyo Institute of Technology, 4259 Nagatsuta, Midori-ku, Yokohama 226-8503, Japan

Corresponding author: Hideo Hosono

E-mail: hosono@lucid.msl.titech.ac.jp

**Abstract**

$LaCo_2B_2$ with a $ThCr_2Si_2$-type structure composed of alternately stacked La and CoB layers exhibits metallic electrical conductivity and Pauli paramagnetic behavior down to 2K. Bulk superconductivity with a $T_c$ of ~4K emerges upon substitution with dopant elements; *i.e.*, isovalent substitution to form $(La_{1-x}Y_x)Co_2B_2$, or aliovalent substitution to form $La(Co_{1-x}Fe_x)_2B_2$. Highly covalent bonding between Co 3d and B 2p levels in the CoB layers, which is caused by the B 2p level being shallower than the Fermi level, removes magnetic ordering from Co 3d electrons even in the undoped samples.

Since the discovery of high temperature superconductivity in F-doped LaFeAsO [1], iron-based layered pnictides including $AFe_2As_2$ (A = Sr, Ba) and FeSe have attracted much attention because superconductivity with a relatively high superconducting transition ($T_c$) was observed from their magnetic iron-based parent compounds. Intensive research achieved a $T_c$ of 57K in $SmFeAs(O_{1-x}F_x)$ [2], which is second only to that of cuprate superconductors [3]. These pnictides or chalcogenides adopt tetragonal crystal structures that include FeAs layers composed of edge-sharing As tetrahedra with an Fe center within the plane. Doping the parent compound with carriers suppresses the formation of antiferromagnetic (AFM) ordering, resulting in the emergence of superconductivity in most of these Fe pnictides, similar to cuprate superconductors. Many Ni-based compounds including $LuNi_2B_2C$, $YNi_2B_2$, $MgCNi_3$, and LaNiPO also exhibit superconductivity [4-6]. These metallic compounds possess Pauli paramagnetic (PM) behavior and exhibit superconducting transitions without the aid of carrier doping. Among these Ni-based compounds, those with a NiB layer, an analog of the FeAs layer in Fe pnictides with respect to structure and the number of 3d electrons, have tendency to exhibit a relatively high $T_c$ (10-15K) [4,5]. For the past 25 years it has been known that some late transition metal compounds show superconductivity at low temperature, as evidenced by these compounds composed of Fe, Ni and Cu. These results suggest that compounds of Co, which is located between Fe and Ni in the periodic table, should exhibit superconducting transitions. Nonetheless, very few reports on superconductive Co-based compounds are available [7,8]; for instance, LaCoPO and LaCoAsO are itinerant ferromagnetic metals but do not exhibit a superconducting transition [9] even though they have the same crystal structure as LaFeAsO.

Here, we report that $LaCo_2B_2$ with a negative Seebeck coefficient exhibits a $T_c$ of ~4K upon cationic substitution. The parent compound, $LaCo_2B_2$, which was first reported by Niihara *et al.* in 1973 [10], adopts a tetragonal $ThCr_2Si_2$-type (so called 122-type) structure, to which more than 700 isostructural compounds including $AFe_2As_2$ (A=Sr or Ba) [11-13] belong. It is expected that Co-based materials will be a platform from which to explore new superconductors in the 122-type family, even though their $T_c$ is only moderately high.

Polycrystalline samples of $(La_{1-x}Y_x)Co_2B_2$, $La(Co_{1-x}Fe_x)_2B_2$, and $LaCo_2(B_{1-x}Si_x)_2$ were synthesized by the arc-melting method. Elemental La, Co, B, Y, Fe and Si were used as starting materials. The arc-melted ingot were annealed at 1273K in an evacuated silica ampoule when necessary. The chemical compositions of the products were determined using a JXA-8530F electron microprobe analyzer (JEOL). The crystal structures of the synthesized materials were examined by powder X-ray diffraction (XRD) at 300K using Cu $K_\alpha$ or synchrotron radiation (BL02B2 beamline of SPring-8) with the aid of Rietveld refinements using RIETAN-FP software [14]. The dependence of DC electrical resistivity ($\rho$) on temperature was measured over 2-300K by a conventional four-probe method. Magnetization (M) measurements were performed with a vibrating sample magnetometer (Quantum Design).

Spin-polarized density functional theory (DFT) periodic calculations were performed using the Vienna Ab-initio Simulation Package code [15] with a projector augmented wave method [16,17] and PBE96 generalized gradient approximation (GGA) functional. The GGA+U approximation was applied to the La 4f orbitals using the rotational-variant +U method with an effective Coulomb parameter ($U_{eff}$) of 11 eV. For Co 3d, different $U_{eff}$ values varied from 0 to 1.2 eV, whose value is proposed for Fe 3d in LaFeAsO[18], were examined. First, quantum-mechanically stable structure was calculated so as to take a SCF energy minimum. Band structure and density of states (DOS) were calculated based on the relaxed structures.

The arc-melted $LaCo_2B_2$ ingots were dark gray in color and possessed a metallic luster.

Powder XRD measurements revealed that the crystal structure was of 122-type (tetragonal lattice, space group *I*4/*mmm*, No. 139), and the chemical composition determined by electron microprobe analysis was $LaCo_2B_2$. The results of crystal structure refinements using synchrotron powder XRD are summarized in **Table I**, while the observed, calculated, and difference patterns are shown in **Fig. 1(a)**. Structure relaxation calculations were performed using DFT. These revealed that the total magnetic moments were always zero even if $U_{eff}$ of Co 3d was varied from 0 to 1.2 eV, corresponding to a non-spin-polarized ground state. The calculated lattice parameters reproduce experimental values with errors of less than 0.3%, and z(Boron) on a 4e site agrees well with those obtained from Rietveld analyses, as summarized in **Table 1**. **Fig. 1(a)** also shows the determined crystal structure of $LaCo_2B_2$. In the structure, CoB and La layers are stacked alternately along the c-axis. Each Co ion is coordinated by four B ions forming a fluorite-type layer, the length of each Co-B bond is 0.20 nm and the B-Co-B angles are 131 and 100°, indicating that the CoB layer is significantly compressed along the c-axis. This structural feature is similar to that observed for the Co-P tetrahedral in LaCoPO with P-Co-P angles of 123 and 103° [9]. In the fluorite-type layer, the Co ion occupies a 4d site with $D_{2d}$ symmetry, resulting in the formation of a Co square lattice net. Note that the Co-Co distance in the net is 0.255 nm (= $a/\sqrt{2}$), which is very close to that in Co with a hexagonal close-packed arrangement (~0.250 nm) [19]. The B ion occupies a 4e site on a 4-fold rotational axis, and resides at the apex of a square pyramid. Anionic B ions often prefer to condense in solids to form B-B bonds with a typical distance of 0.16-0.19 nm, as observed in FeB, $MgB_2$, and $CaB_6$ [20]. However, the obtained B-B distance of 0.34 nm between the CoB layers in $LaCo_2B_2$ is far longer than that expected from the covalent radius of B (0.141 nm), indicating that interlayer B-B bonding is not present. It is likely that the large size of the La ion sandwiched by the CoB layers prevents dimerization of B ions. Borides with a 122-type structure are quite rare except for Co borides [5,10], making this structure particularly unusual. It should also be noted that the La–B distance (0.308 nm) is significantly larger than their ionic or atomic radius (0.116 nm for $La^{+3}$) [21].

Isovalent doped compounds, $(La_{1-x}Y_x)Co_2B_2$, hole-doped compounds, $La(Co_{1-x}Fe_x)_2B_2$, and an electron-doped compound, $LaCo_2(B_{1-x}Si_x)_2$, were also synthesized. The unitcell volume and lattice constants change monotonically with the ionic radii of the dopants (data not shown), indicating the formation of a solid solution up to x = 0.20, 0.30, and 0.10 for Y, Fe, and Si, respectively.

**Figure 2(a)** shows the resistivity-temperature (ρ-T) curve for $LaCo_2B_2$ and $(La_{1-x}Y_x)Co_2B_2$ (x = 0, 0.10) under an applied magnetic field of $H$ = 0 Oe. The resistivity for $LaCo_2B_2$ (x = 0), which is a good metal, is $\sim 7 \times 10^{-5}$ Ω cm at 300K, and it shows a small drop at 4K. The Seebeck coefficient at 300K is -7.4 μV/K, indicating that the electron carriers are responsible for conduction. No distinct peaks originating from magnetic transitions were observed down to 2K in magnetic susceptibility measurements (data not shown), which indicates a Pauli PM state and is consistent with the calculated ground state. Substitution of Y into 10% of the La sites increases ρ, while the structure of the CoB conduction layer remains unchanged. The resistivity of the 10% Y-doped $LaCo_2B_2$ decreases almost linearly with T for T > 130K. At T < 20K, ρ shows a dependence of $T^2$, indicating a Fermi liquid state. As shown in the inset of **Fig. 2(a)**, a sharp drop in ρ was observed at T = 4.4K, and disappeared at 4.1K under zero magnetic field ($H$ = 0 Oe). The zero-resistivity temperature decreases with increasing $H$, suggesting that 10% Y-doped $LaCo_2B_2$ undergoes a superconducting transition at 4.1K. **Fig. 2(b)** shows the temperature dependence of the magnetic susceptibility (χ) of 10% Y-doped $LaCo_2B_2$ measured in a zero-field cooling (ZFC) process and a field cooling (FC) process at 10 Oe. Between 4.2 and 300K, χ was very small and nearly independent of temperature, implying Pauli PM. At 4.2K, χ began to decrease, becoming negative and reaching -0.24 emu/mol-Co. This value corresponds to a shielding volume fraction of 15% (estimated from the χ value for perfect diamagnetism) at 2K, which confirms that the bulk superconductivity transition occurs at

4.2K. The M-H curve at 2K presented in the inset of **Fig. 2(b)** shows a typical profile for a type-II superconductor with a lower superconducting critical magnetic field ($H_{c1}$) of ~90 Oe. It was confirmed that $T_c$ did not depend on x. The observed phenomena are similar to the chemical pressure-driven superconductivity reported for $BaFe_2(As_{1-x}P_x)_2$ [12].

**Figure 3** shows ρ-T curves for $La(Co_{1-x}Fe_x)_2B_2$ (x = 0-0.30). Co doping of x ≥ 0.10 induced superconductivity transitions at $T_c$ ~ 4K. Substitution of Fe into Co sites is expected to result in hole doping but the resistivity increased, which is explained by enhanced impurity scattering caused by direct doping of impurities into the CoB conduction layer. This is in sharp contrast to Co-doped $BaFe_2As_2$, where direct doping of Co in the FeAs conduction layer induces superconductivity but it is expected to be electron doping [13]. On the other hand, substitution of Si into B sites did not induce a superconducting transition above 2K.

The electronic nature of $LaCo_2B_2$ was investigated computationally as well as experimentally. **Figure 1(b)** shows partial density of states (PDOS) for $LaCo_2B_2$ calculated by DFT. The three following factors are evident; (1) La 4f and 5d levels are located >5 eV higher than the Fermi energy ($E_F$), (2) $E_F$ is composed mainly of Co 3d and B 2p orbitals, and (3) PDOS of Co 3d and B 2p orbitals were similar to each other. (1) shows that the +3 charge state is the most stable for La in this compound because any introduced electrons would occupy the 4f and/or 5d levels that are located at much higher energy than $E_F$. A straightforward conclusion from (2) and (3) is that metallic conduction occurs in the CoB layer composed of highly covalent Co and B. Here, it is noted that B 2p levels are rather shallow compared with analogous 122-type pnictide compounds (As, or P) in which As 4p and P 3p levels are much deeper than the transition metal 3d levels that primarily occupy $E_F$. The relatively shallow energy of anionic B 2p orbitals makes it difficult for it to adopt a closed shell state ($2p^6$) without the aid of a strongly positive cation. The shallow B 2p orbitals are not powerful enough to fully oxidize a Co ion. As a result, strong Co 3d-B 2p covalent bonding forms a band with a width of 7.1 eV, removing the spin moment of the Co 3d electrons. This is in contrast to the ferromagnetic metal compounds LaCoXO (X=P or As), in which the formal charge of the Co ion can be regarded as +2 by assuming that of X to be -3 [9]. It is likely that the difference originates from the relatively deep energy of the As 4p level compared with that of B 2p, and is similar to the relationship between LaFeAsO and LaFePO, where Fe 3d bands hybridized by broader P 3p and narrower As 4p levels [22] give rise to a PM state and AFM ordering, respectively.

The evolution of superconductivity caused by isovalent (Y) or aliovalent (Fe) substitution suggests that the superconducting transition is not necessary triggered by modification of $E_F$ (or carrier concentration), but that a small structural perturbation caused by the formation of an isovalent solid solution can also induce a superconducting transition. The effect of annealing on the superconductivity of undoped $LaCo_2B_2$ also shows that a small perturbation like improved crystallinity can affect $T_c$. This situation is in contrast with that of FeAs-based compounds where the DOS at $E_F$ is primarily composed of Fe 3d orbitals, and suppression of a spin-density-wave state by aliovalent substitution allows a superconducting phase to form [1,11,13], although the experimental results are similar to those of doped $BaFe_2As_2$ at first sight. While Co is located between Fe and Ni in the periodic table, the electronic nature of $LaCo_2B_2$ with Pauli PM behavior caused by strong covalent bonding in transition metal borides appears to resemble NiB-based compounds.

The electrical and magnetic properties of $LaCo_2X_2$ (X = P, Si, or B) at low temperature are changed significantly by varying X from P to B: the materials are ferromagnetic with $T_{Curie}$ = 103K for P [23], PM metallic conductor for Si [24], and superconductor for B. Anionic substitution of the parent $LaCo_2B_2$ with P or Si corresponds to electron doping. A variety of derivatives can be synthesized by doping La, Co, or X sites, which allows for the systematic study of superconductivity.

In summary, it was found that LaCo$_2$B$_2$ composed of alternate stacked layers of La and CoB exhibits bulk superconductivity with a T$_c$ of ~4K upon cationic substitution. The appearance of superconductivity by isovalent or aliovalent substitution suggests that the highly covalent CoB layer exhibits superconductivity by introduction of a small change, such as structural randomness caused by the formation of a solid solution.

This work was supported by the Funding Program for World-Leading Innovative R&D on Science and Technology (FIRST), Japan. We thank Dr. J. Kim (JASRI), Dr. S. W. Kim and Dr. H. Hiramatsu (Tokyo Tech) for XRD measurements.

**Table I.** Crystal structure of LaCo$_2$B$_2$,[a] as determined by Rietveld refinement[b] of a synchrotron XRD data set covering a 2θ range of 3-70°. The wavelength was resolved to 0.35290(1) Å against a CeO$_2$ standard. Data for the crystal structure obtained from DFT calculations is also shown.

| | a (nm) | c (nm) | z(Boron) | B(La)(Å$^2$) | B(Co) (Å$^2$) | B(Boron) (Å$^2$) |
|---|---|---|---|---|---|---|
| Obs.(300K) | 0.361080(9) | 1.02051(2) | 0.3310(5) | 0.42(1) | 0.10(1) | 2.4(1) |
| Calcd. | 0.3606 | 1.0232 | 0.333 | - | - | - |

[a]Wyckoff positions for the ThCr$_2$Si$_2$−type structure (space group: *I*4/*mmm* (No. 139)) are as follows: B at the 4e site (0, 0, z(Boron)); Co at the 4d site (1/2, 0, 1/4); and La at the 2a site (0, 0, 0). [b]R$_{wp}$ = 4.86%, R$_I$ = 1.33%, and goodness of fit S = 1.9.

# Figure Captions

**FIG. 1. (a)** (Bottom) Observed synchrotron powder XRD pattern and results from Rietveld refinement for $LaCo_2B_2$. XRD data were collected by a Debye-Scherrer camera. The dots and lines indicate the observed and calculated patterns, respectively. The difference between the observed and calculated patterns is shown at the bottom. The vertical marks indicate the Bragg reflection positions for $LaCo_2B_2$. (Top) Crystal structure of $LaCo_2B_2$. **(b)** Calculated DOS of $LaCo_2B_2$, with the PDOS for La, Co, and B.

**FIG. 2. (a)** $\rho$-T plots for $(La_{1-x}Y_x)Co_2B_2$. The inset shows $\rho$-T curves for the sample with x=0.10 as a function of magnetic field. **(b)** Temperature dependence of the magnetic susceptibility ($\chi$) of $(La_{0.9}Y_{0.1})Co_2B_2$ under ZFC and FC conditions at 10 Oe. The inset shows the field dependence of magnetization at 2K.

**FIG. 3.** $\rho$-T plots for $La(Co_{1-x}Fe_x)_2B_2$ (x = 0.0-0.3).

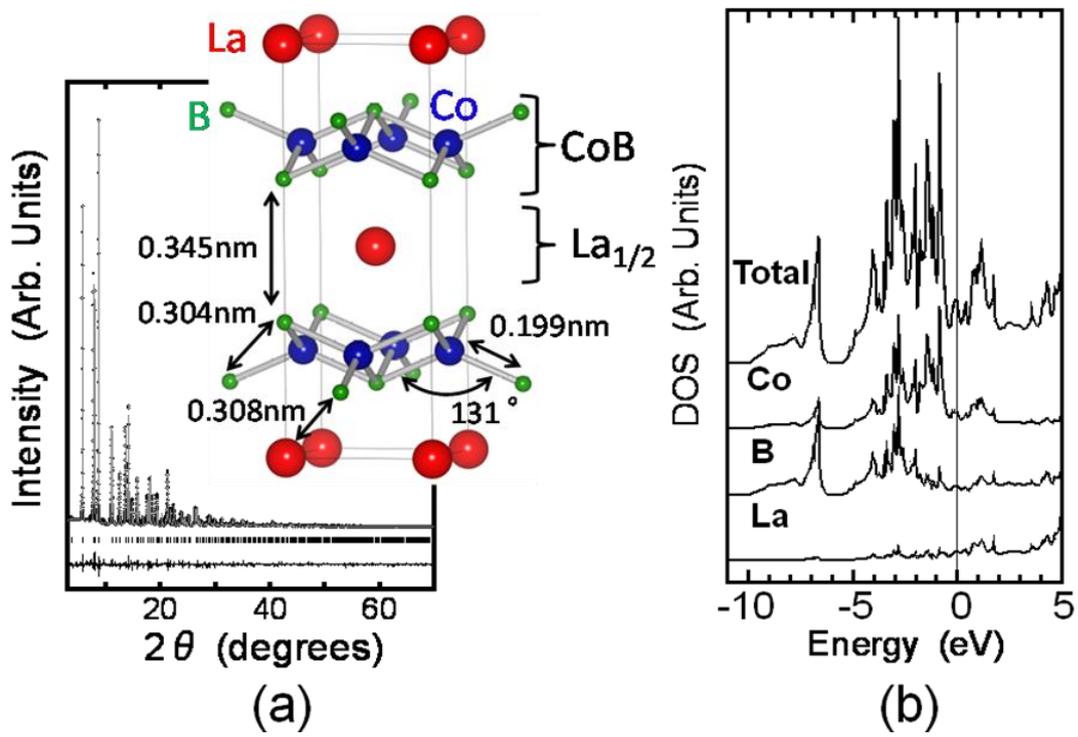

**FIG. 1**

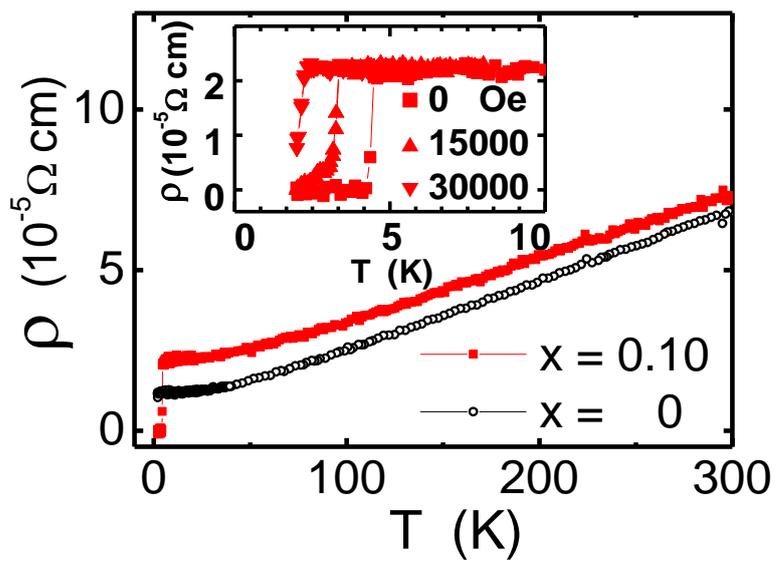

**FIG. 2(a).**

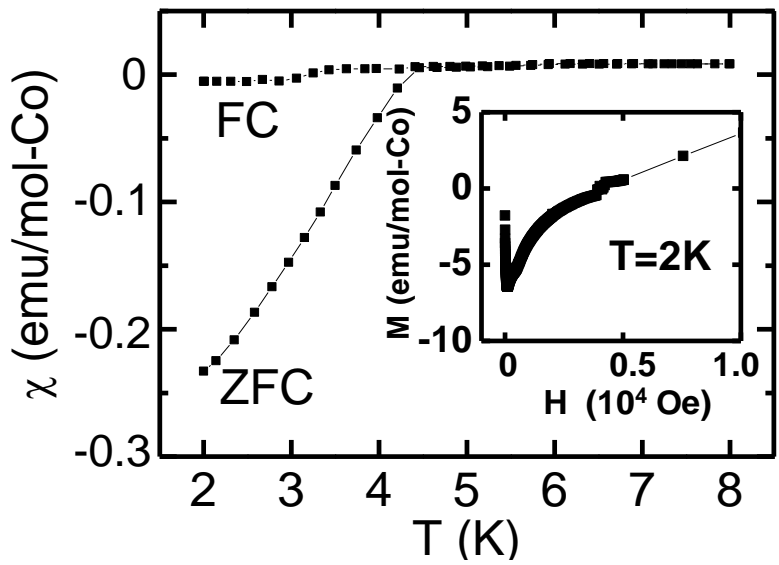

**FIG. 2(b)**

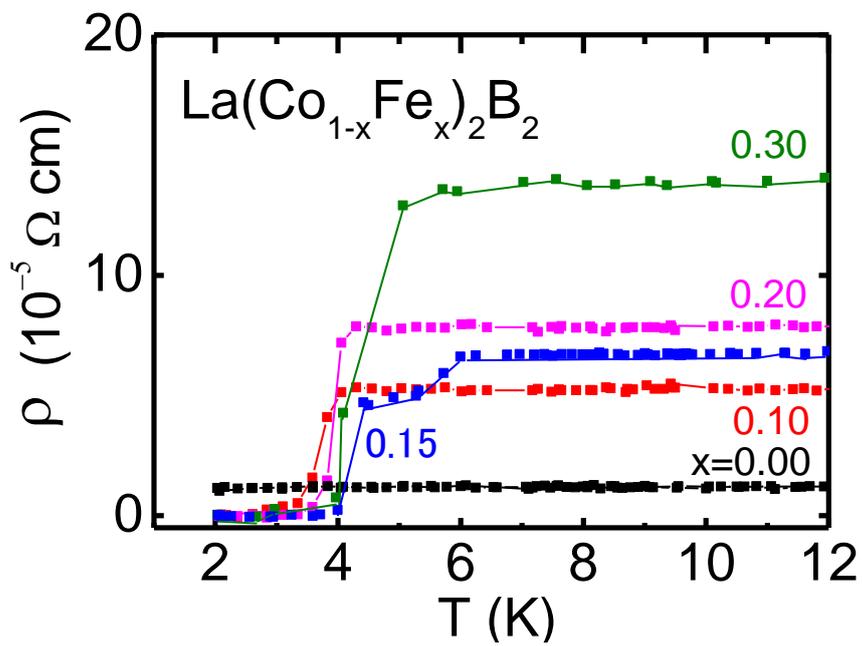

**FIG. 3**